# Complex Morlet Wavelet Analysis of the DNA Frequency Chaos Game Signal and Revealing Specific Motifs of Introns in C.elegans

Imen Messaoudi[1], Afef Elloumi Oueslati[1] and Zied Lachiri[2]

[1]*Université de Tunis El Manar, Ecole Nationale d'Ingénieurs de Tunis, LR-11-ES17*
*Signal, Images et Technologies de l'Information, BP 37, le Belvédère, 1002, Tunis, Tunisie.*

[1]imen_messaoudi@ymail.com

[1]afefelloumi@gmail.com

[2]*Département de Génie Physique et Instrumentation,*
*INSAT, BP 676, Centre Urbain Cedex, 1080, Tunis, Tunisie.*

[2]Zied.lachiri@enit.rnu.tn

***Abstract*— Nowadays, studying introns is becoming a very promising field in the genomics. Even though they play a role in the dynamic regulation of gene and in the organism's evolution, introns have not attracted enough attention like exons did; especially of digital signal processing researchers. Thus, we focus on analysis of the *C.elegans* introns. In this paper, we propose the complex Morlet wavelet analysis to investigate introns' characterization in the *C.elegans* genes. However, catching the change in frequency response with respect to time of the gene sequences is hindered by their presence in the form of strings of characters. This can only be counteracted by assigning numerical values to each of the DNA characters. This operation defines the so called "DNA coding approach". In this context, we propose a new coding technique based on the Frequency Chaos Game Representation (FCGR) that we name the "Frequency Chaos Game Signal" (FCGS).**

**Results of the complex Morlet wavelet Analysis applied to the *Celegans* FCGS are showing a very distinguished texture. The visual interpretation of the colour scalograms is proved to be an efficient tool for revealing significant information about intronic sequences.**

***Keywords*— Complex Morlet wavelet Transform, Frequency Chaos Game Signal, Introns, motifs, *Caenorhabditis elegans*.**

## I. INTRODUCTION

Actually, genes analysis is becoming a major topic because of their implication in a large number of diseases, mainly those involved in the mechanisms of heredity. Knowing the gene location, composition and structure is then needed in order to identify the genetic component which is responsible for the considered disease. Such information can be helpful also in understanding the dynamic regulation of gene expression which determines, in addition, the organism's complexity. It can be useful, in the other hand, in studying genes evolution and hence the organisms evolution.

Furthermore, it is important to notice that the current model of the living organisms divide them into two main classes: eukaryotes and prokaryotes. In prokaryotes, genes consist only of coding regions. In the contrary, in eukaryotes the coding regions which are termed "exons" are interrupted by non-coding zones which are called "introns". Note that most eukayotic gene sequences are broken up by one or more introns, with a great variation in size.

Due to the importance of genes, several experimental and computational methods are developed in recent years aiming at providing accurate genomes annotations. Thus, localizing DNA that encodes genes was of prime importance in the genomic field. The process of identifying these zones is commonly called gene prediction.

Typically, the gene prediction processes are focused on detecting the coding areas in species' genomes. The key factor determining the orientation of researches on coding DNA prediction turns on that exons are characterized by a periodic organization of 3 base pairs (bp). This property has been at the core of several methods of exons prediction. In this context, Signal processing tools have played a determinant role in handling genomic data [15], [16], [17], [24] especially in establishing the exon prediction approaches and the gene annotations [22].

The identification of the protein coding areas was essentially based on the Discrete Fourier transform (DFT) [13], [14], [18], [19], [20], [22]. Even if the spectrum analysis (based on DFT) is the most commonly used method for exons prediction, it presents an obvious weakness. In fact, the DFT analysis introduces spectral perturbation due to the fixed length of the analysis window. The type of the window as well as its width must be chosen in such way that they provide good frequency resolution. This is a difficult task because that genes are neither continuous nor contiguous and have a variable size.

To resolve this problem, other studies focused on the digital filters [7], [12]. Such approach is shown to be efficient but it led to results which are comparable to those of the spectral analysis [11]. Thus, a relevant tool permitting an accurate prediction is needed; such is the case of the Wavelet analysis. In [10], a modification to the Gabor wavelet transform was introduced (MGWT). Aiming at localizing the 3 bp periodicity, the MGWT algorithm is based on fixing the wavelet basic frequency at 1/3 independently of the scale parameter.

Globally, these techniques rely on the exons prediction based on the property of the 3 bp periodicity. But till now, approaches for characterizing and identifying the non-coding portions of eukaryotic genes, have not yet emanated. In fact, although introns are non-functional DNA, they play an important role in the organism's evolution. Indeed, they form an easy target to mutations over time. Thus, studying introns in the genomes is a promising direction in the genomics.

In this paper, we focus on the intronic sequences analysis in the *Caenorhabditis elegans* genome (*C.elegans*) as part of the genomic signal processing discipline. In this sense, we propose the complex Morlet wavelet analysis to identify the location of the non-coding regions in the Y50D7A.4 gene. To reach this goal, the conversion of the DNA data set into a numerical sequence is imperative. So, we provide a new technique for DNA representation based on the Frequency Chaos Game Representation (FCGR). This computational and visual tool is named the Frequency Chaos Game Signal (FCGS). The efficiency of our coding method in the intron characterization will be then proved by interpreting the resulting scalograms.

This paper consists of five sections. Section 2 introduces the Frequency Chaos Game Signal and exposes the main steps required to generate a numerical DNA representation. Section 3 presents an overview on the Continuous Wavelet Transform, as well as the Complex Morlet wavelet which is taken as the mother wavelet function. Section 4 deals with the exploration of the intronic sequences into the *C.elegans* genome by interpreting the Morlet wavelet scalograms. Finally, section 5 concludes the content of this paper.

## II. THE FREQUENCY CHAOS GAME SIGNAL: A NEW DNA CODING TECHNIQUE

The DNA coding is a primordial step that precedes any Digital Signal Processing based applications. The basic principle consists on assigning numerical value to each of the DNA characters (Adenine 'A', Cytosine 'C', Guanine 'G' and Thymine 'T') or of considered character group (for example 'AA', 'CC', 'CA', etc).

In this section we present a new coding technique based on the Frequency Chaos Game Representation (FCGR); which is the Frequency Chaos Game Signal (FCGS). Our method is original in that it allows us to follow the frequency evolution of words' occurrence along a given sequence, regardless of the words' size. In other word, we provide a temporal version of the FCGR which is basically a two Dimensional (2D) representation. For this fact, it is required to pass through the Frequency Chaos Game Representation.

The FCGR technique is submerged from the Chaos Game Representation theory (CGR), which permit in turn the representation of bio-molecular sequence into a 2D plot.

The CGR is an iterative algorithm consisting on mapping a DNA sequence into a unit square in such way that we obtain a scatter dots plot. The four nucleotides A, C, G and T are then placed at the corners as described in figure 1, [1], [2].

If we consider the sequence $U$={$U_1$,$U_2$, …$U_N$}, the first point $X_0$ is placed at the center of the square. The next point $X_{n+1}$ is repeatedly placed halfway between the previous plotted point $X_n$ and the segment joining the vertex corresponding to the read letter $U_{n+1}$ [1], [2], [9]. The coordinates of each plotted point are given by:

$$X_{n+1} = \frac{1}{2}\left(x_n + \ell_{U_{n+1}}\right) \quad (1)$$

Where $\ell_{Un+1}$ can be:

$$\ell_A(0,0),\ \ell_C(0,1),\ \ell_G(1,1) \text{ or } \ell_T(1,0) \quad (2)$$

The figure 1 illustrates the procedure to represent the sequence "ACGGT".

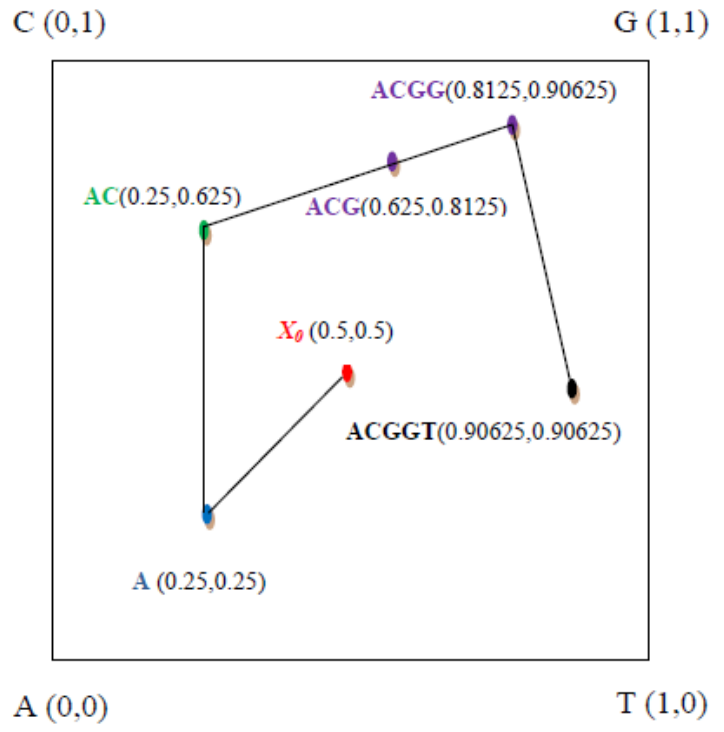

Fig. 1 Representation of the sequence "ACGGT" by the CGR technique: the word "A" is represented by the dot $X_1$=1/2($X_0$+ $l_{U1}$)=1/2((0.5,0.5)+(0,0)). Likewise, the word "AC" is represented by the point $X_2$, after that the word "ACG" is represented by $X_3$, then the word "ACGG" is represented by $X_4$ and finally the word "ACGGT" is represented by $X_5$. The set of points {$X_1$, $X_2$, $X_3$, $X_4$ and $X_5$} forms the final CGR plot.

To move towards the Frequency Chaos Game Representation, the CGR image must be divided into $4^k$ sub-squares. Each sub-square is, in turn, associated to a *k*-lengthen sub-pattern according to a specific organization. The frequency of a *k*-lengthen word occurrence is equal to the number of counted dots in the correspondent sub-square, divided by the complete length of the DNA sequence. The frequencies' set of all possible words consisting of *k* bases form the elements of the FCGR matrix. Each frequency value in the FCGR matrix is then coded according to a color scale. In figure 2, we furnish an example of the Frequency Chaos Game Representations with order {1, 2 and 3} as well as the arrangement of all possible words into the FCGR's sub-squares.

Note that the darkest pixels represent the highest frequency values, whereas the clearest ones represent the lowest frequencies [3], [8].

Within the framework of DNA coding, we propose a coding measure scheme employing the Frequency Chaos Game matrices. The new coding approach is named the Frequency Chaos Game Signal (FCGS). The FCGS technique is in the spirit of temporizing the FCGR representation and consists of a one dimensional signal. The basic principle is that we attribute the frequency value of each sub-pattern to the same group of nucleotides existing in the sequence. Thus, we have to follow the following steps:

1. If we consider words consisting on $n$ nucleotides, we have to generate the $n^{th}$-order FCGR for the whole sequence ($FCGR_n$).
2. We extract the first $n$-lengthen word which is positioned in the beginning of the sequence.
3. We extract the frequency value of the read word from the $FCGR_n$ matrix.
4. We attribute, to the first position, the frequency value. Then, we move to position 2.
5. We redo the same procedure till reaching the position: Pos= $L$-$n$+1, where $L$ is the DNA sequence length and $n$ the word's length.

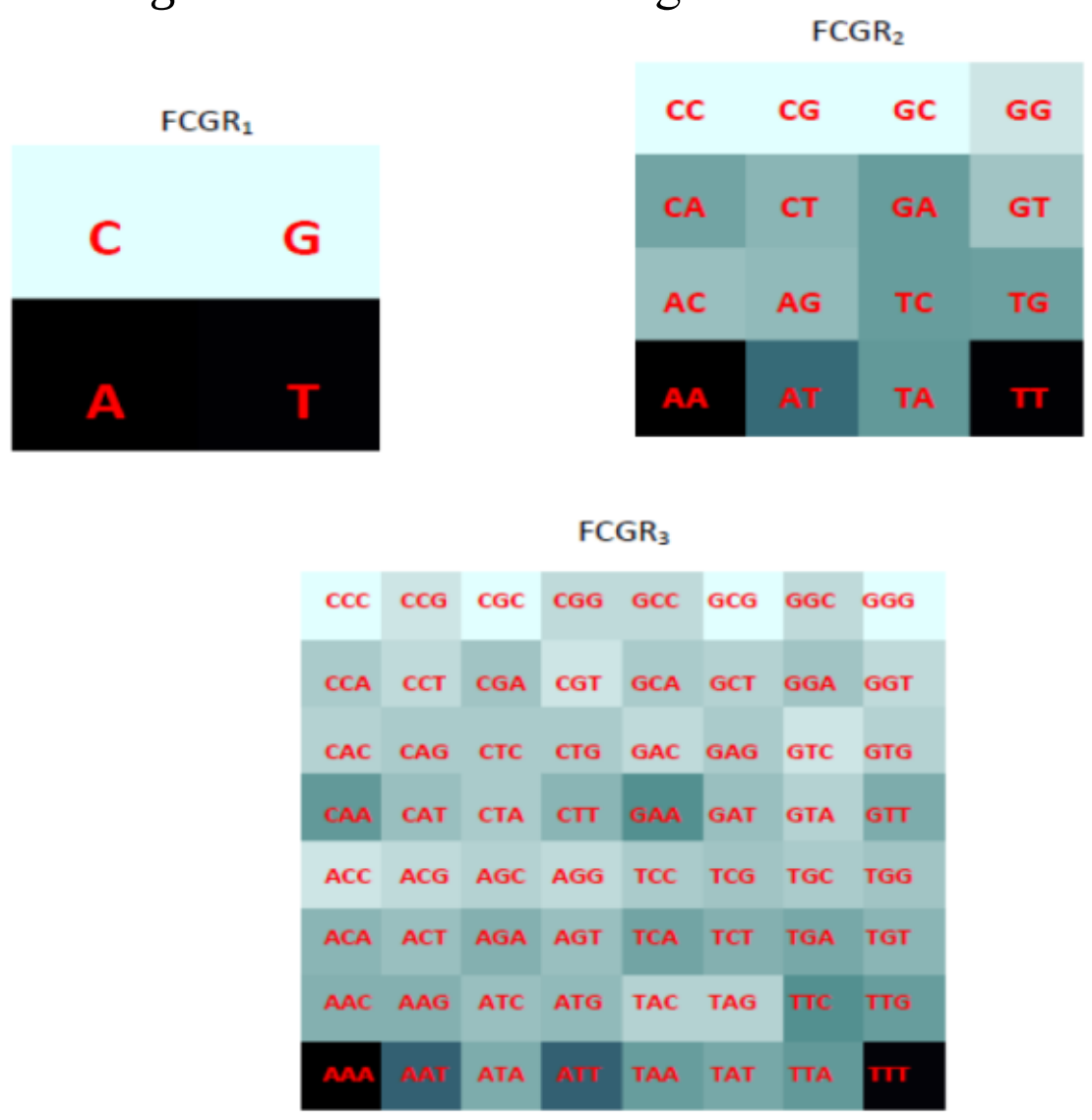

Fig. 2 $FCGR_1$, $FCGR_2$ and $FCGR_3$ of the *C.elegans* chromosome V and the distribution of the possible words having the length $k$={1,2 and 3} in the Frequency Chaos Game Representation' s space.

We consider, for example, a sequence $S$ taken from the chromosome V of *C.elegans*; $S$={ GAATTCCTAAGCCTAAGCCT}. The motifs contained in the sequence $S$ with size one, two and three are as follows:
Monomers={ G,A,A,T,T,C,C,T,A,A,G,C,C,T,A,A,G,C,C,T}.
Dimers={ GA,AA,AT,TT,TC, CC, CT, TA, AA, AG, GC, CC, CT, TA, AA, AG, GC, CC, CT}.
Trimers={ GAA,AAT,ATT,TTC,TCC,CCT,CTA,TAA,AAG, AGC,GCC,CCT,CTA,TAA,AAG,AGC,GCC,CCT }.

While the sequence $S$ belongs to the chromosome V, we must compute $FCGR_1$, $FCGR_2$ and $FCGR_3$ for the considered chromosome. The resulting matrices are shown below. Based on these frequencies, we assign the correspondent value to each of the monomers, dimers and trimers previously extracted. In figure 3, we present the obtained signals: $FCGS_1$, $FCGS_2$ and $FCGS_3$.

$$FCGR_1 = \begin{pmatrix} 0.1774 & 0.1769 \\ 0.3226 & 0.3231 \end{pmatrix}, FCGR_2 = \begin{pmatrix} 0.0333 & 0.0305 & 0.0333 & 0.033 \\ 0.0627 & 0.0509 & 0.0618 & 0.0487 \\ 0.0489 & 0.0506 & 0.0619 & 0.0628 \\ 0.1339 & 0.0893 & 0.0642 & 0.1341 \end{pmatrix}$$

$$FCGR_3 = \begin{pmatrix} 0.0057 & 0.0068 & 0.0055 & 0.0067 & 0.0069 & 0.0055 & 0.0068 & 0.0057 \\ 0.0125 & 0.0083 & 0.0106 & 0.0078 & 0.0113 & 0.0097 & 0.012 & 0.0085 \\ 0.0106 & 0.0116 & 0.0112 & 0.0117 & 0.0087 & 0.0111 & 0.0086 & 0.0105 \\ 0.0244 & 0.0161 & 0.0107 & 0.0174 & 0.0267 & 0.0153 & 0.0112 & 0.0183 \\ 0.0085 & 0.0078 & 0.0096 & 0.0081 & 0.0121 & 0.0106 & 0.0114 & 0.0125 \\ 0.0175 & 0.0151 & 0.0178 & 0.015 & 0.0214 & 0.0178 & 0.0214 & 0.0175 \\ 0.0182 & 0.0172 & 0.0154 & 0.0161 & 0.0113 & 0.0106 & 0.0268 & 0.0245 \\ 0.0611 & 0.0373 & 0.0205 & 0.0373 & 0.0217 & 0.0206 & 0.0218 & 0.0611 \end{pmatrix}$$

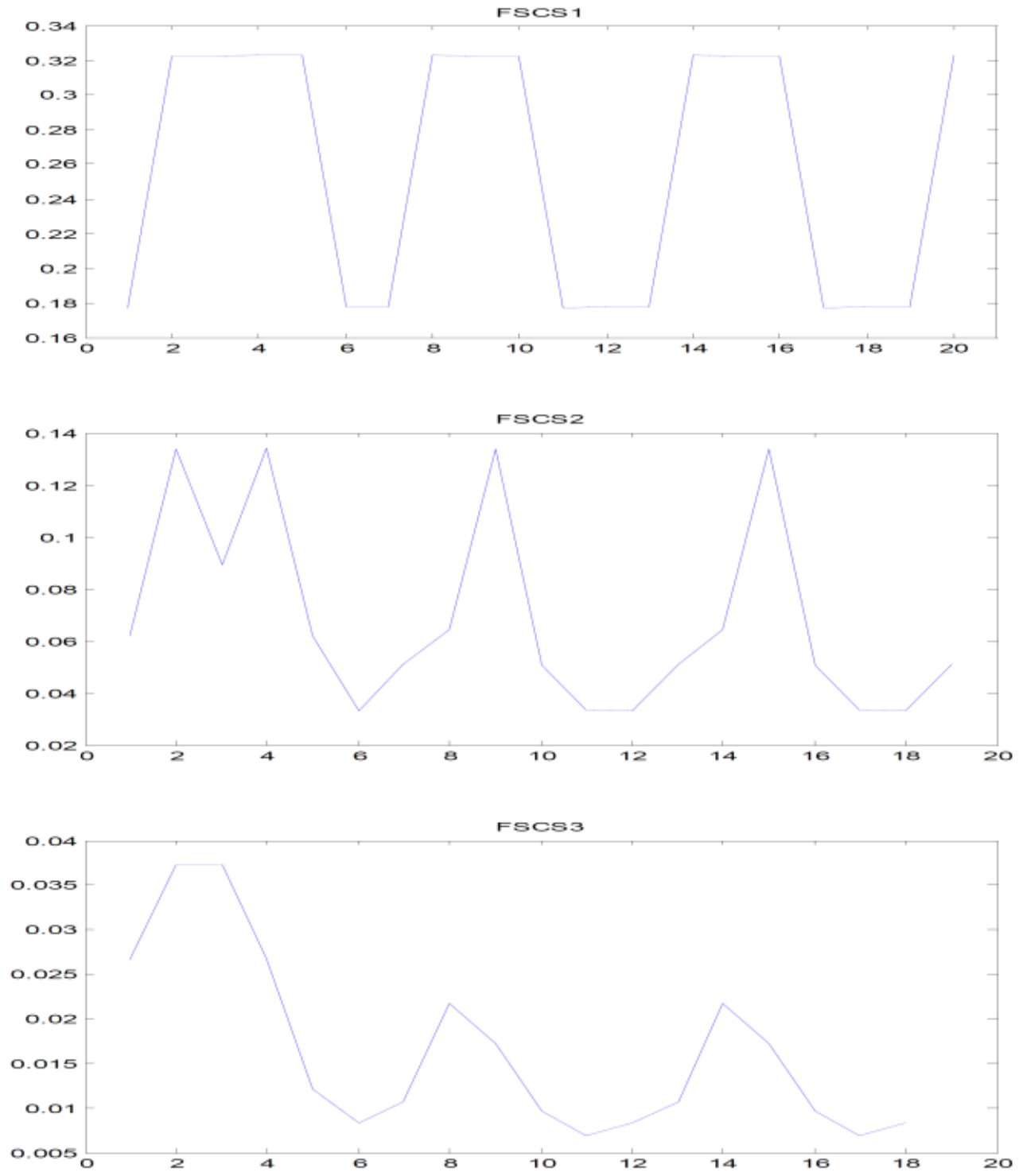

Fig. 3 $FCGS_1$, $FCGS_2$ and $FCGS_3$ plots of the sequence S={GAATTCCTAAGCCTAAGCCT}which is taken from the *C.elegans* chromosome V.

The advantage of our technique is that we provide different signals for the same sequence depending on a desired scale $n$ (here $n$ is the length of a nucleotides group). This can offer a variation of the information along the bio-molecular sequences. The particularity of our method resides in the fact that we are based on the statistical properties of the genome itself; which may strongly reflect interesting information.
To demonstrate the effectiveness and usefulness of our coding, we chose to apply the complex Morlet wavelet analysis.

## III. The Complex Morlet Wavelet Analysis

A main difficulty that faces researchers in the field of genomics is to find relevant tools and techniques to accurate prediction and detection of DNA hotspots. These regions are of different functions, structure and size, which render their prediction a very hard task. Almost, traditional methods are based on the average of DNA base contents within a fixed window such the case of the Fourier transform based algorithms. However, fixing the window length impedes revealing complete and accurate information. The wavelet analysis appears as an excellent solution to this problem.

The wavelet transform (WT) was introduced by Morlet in 1983 to study seismic signals [5]. Then, the proposed processing was well formalized in 1984 with collaboration of Grossman [5]. Since then, the wavelet theory has been the subject of diverse theoretical developments and practical applications including the bio-molecular ones.

Unlike the Fourier transform, the wavelet transform performs a scale-space representation of the signal; which offers very good time and frequency localization. The analysis procedure consists in decomposing a given signal into a sum of basic functions obtained by translations and expansions of a specific small wave called the mother wavelet. The latter has only a time domain representation as the wavelet function $\psi(t)$ which is a wave-like oscillation. Sets of daughter wavelets are generated by the following expression:

$$\psi_{a,b}(t) = \frac{1}{\sqrt{a}}\psi\left(\frac{t-b}{a}\right), \ a \succ 0, \ b \in \mathbb{R} \tag{3}$$

Where b and a are respectively the time and scale parameters. Assuming that the basic wavelet is positioned around a central frequency $f_0$[1], the frequency set is proportional to scale one. Since, the wavelet transform also leads to a time-frequency analysis [5, 6]. The continuous wavelet transform (CWT) of a function $f(t)$ is defined as:

$$T_{\psi}(f)(a,b) = \frac{1}{\sqrt{a}}\int_{-\infty}^{\infty} f(t)\psi^{*}\left(\frac{t-b}{a}\right) dt \tag{4}$$

Here * represents the operation of complex conjugate. In our analysis, the Complex Morlet wavelet is taken as the wavelet function $\psi(t)$. The Complex Morlet wavelet is a Gaussian-windowed complex sinusoid; its mathematical formulation is given by:

$$\psi(t) = \pi^{-\frac{1}{4}}\left(e^{i\omega_0 t} - e^{-\frac{1}{2}\omega_0^2}\right)e^{-\frac{1}{2}t^2} \tag{5}$$

Where $w_0$ corresponds to the number of oscillations of the wavelet knowing that $w_0$=2*pi* $f_0$. To satisfy the admissibility condition, $w_0$ must be greater than 5. This criterion ensures the mother wavelet to be invertible; which is required by the CWT [4]. Equation (4) gives the so-called wavelet coefficients. The modulus of these coefficients: $|T_{\psi}(a,b)|$ is named scalogram. A simple interpretation of scalograms, allows easy detection of specific DNA regions as well as their underlying feature patterns.

## IV. Experimental results

In this section, we focus on the analysis of the Y50D7A.4 gene, which is located on chromosome III of *C.elegans* (position: [235841:253375]). The considered gene consists of 13 exons and 11 introns. The DNA sequence, as well as the gene annotations, were obtained from the NCBI database [23]. The DNA data set was coded by $FCGS_2$ based on the $FCGR_2$ matrix of the whole chromosome. Once the chromosome sequence was converted to a signal, we computed the continuous wavelet analysis. As for the wavelet parameters, a complex Morlet wavelet which is defined on a temporal support of 600 points was used. In addition, the continuous wavelet transform was performed along scales from 1 to 64 with the parameter $w_0$ fixed at 5.4285. At the aim to explore the content of the resulting scalogram we proceed by a simple zooming. Thus, looking into a window of $10^3$ bp is shown to be sufficient to observe significant information. When we took a closer look at the Y50D7A.4 gene position, we observed the multi-scale spectrum presented in the figure 4.

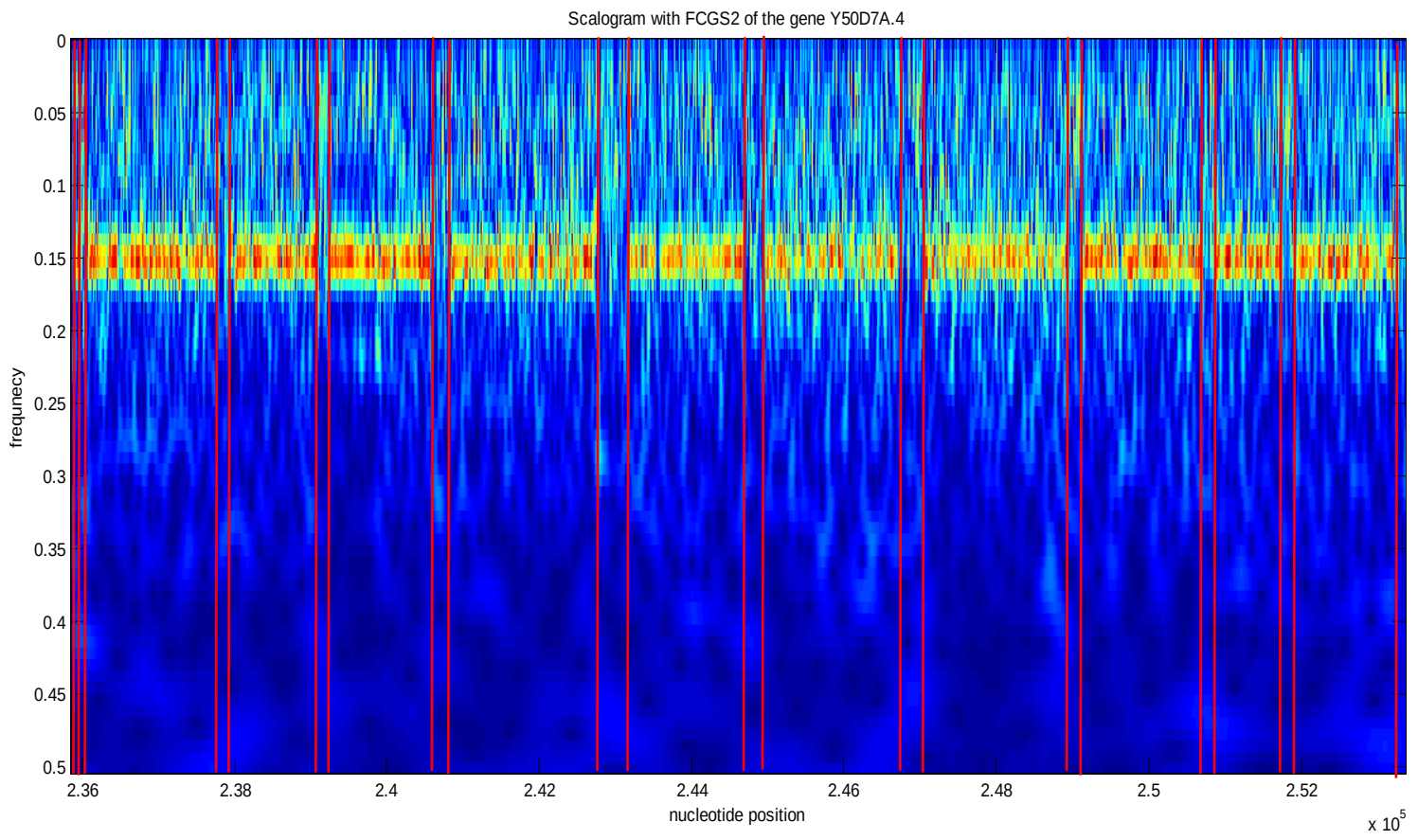

Fig. 4 Scalogram of the Y50D7A.4 gene with the $FCGS_2$ coding technique. The red lines delimit the non-coding regions from the coding ones.

As it can be seen from figure 4, the $FCGS_2$ coding enables the introns characterization instead of exons. The latter appear as blue zones that interrupt a long band with high energy which is spread over the gene totality. This highly band energy is situated around the frequency value of 0.15, which is equivalent to the 6.5 bp periodicity. The available studies concerning genes, don't provide any work that states the characterization of introns by such periodicity. The only work that described the intronic areas as sequences having 6.5 bp periodic patterns is [21]. However, this information is just mentioned as a remark and it is not proved. In our work, we succeeded in proving that introns are characterized by a specific texture that contains periodic motifs of 6.5 bp long.

If we consider the behavior of each of the Y50D7A.4 introns separately, we have to observe the related scalograms which are exposed in figure 5. From the presented sub-figures, we note that they share a common feature (which is the high energy band around the frequency 1/6.5) independently of their lengths. In fact, even the shortest intron (53 bp) is shown to present the frequency 1/6.5. The remaining introns are relatively long and include the same periodic motif at the 0.15 frequency level. A natural way to locate the introns boundaries is to project the wavelet coefficients on the horizontal axis which indicates the nucleotide base position.

Overall, a simple inspection of the scalogram representation, allowed us to catch the exact boundaries between noncoding and coding regions in the *C.elegans* genes.

It is important to note that the available eukaryotic gene prediction tools are not accurate and introduce a high rate of errors [25]. Effectively, we are witnessing about 40% of correct predicted genes [22].

[1] $f_0$ is simply the center of gravity or just the maximum value of the mother wavelet's spectrum.

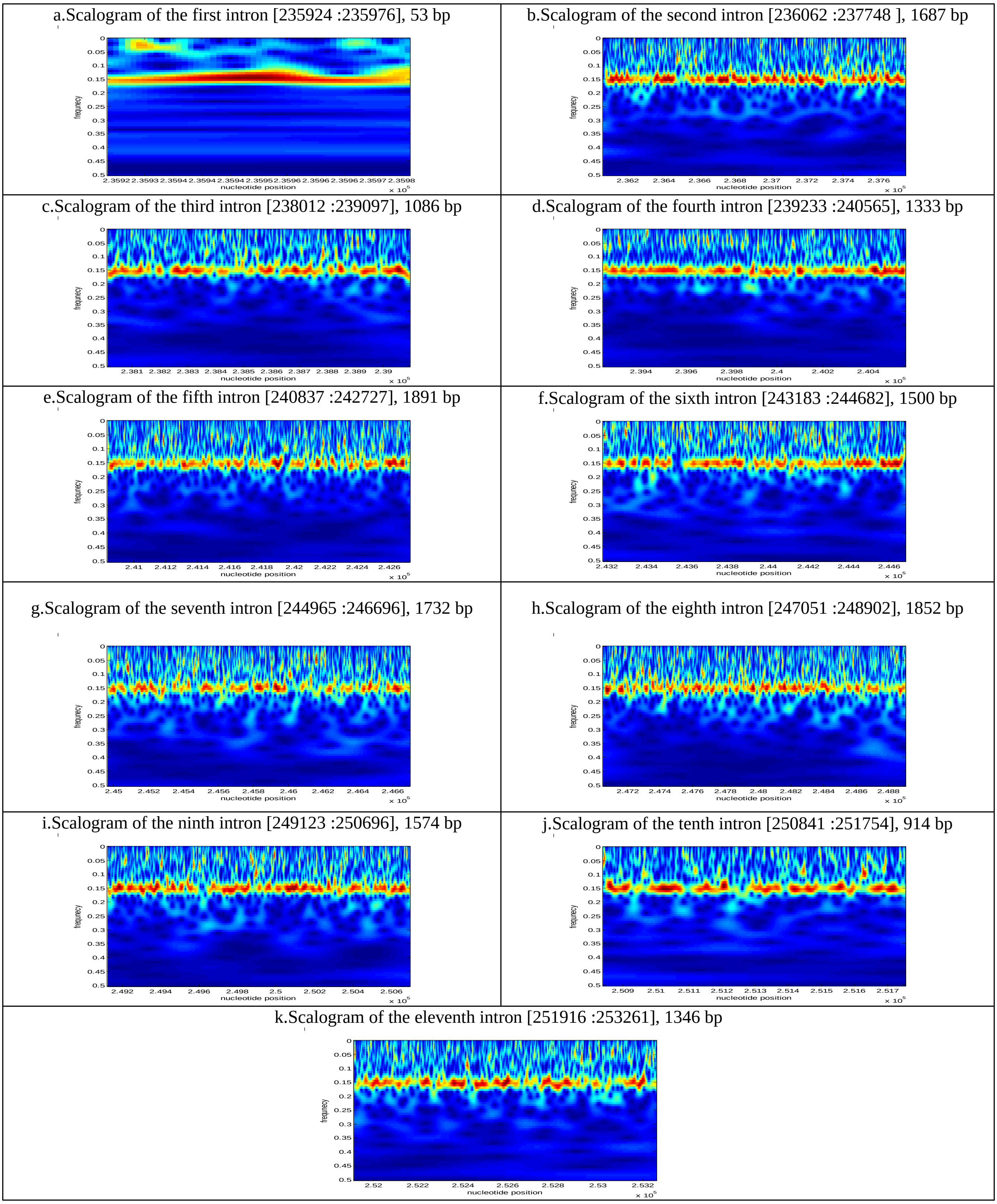

Fig. 5 Scalograms related to the eleven introns present on the Y50D7A.4 gene which are coded with the $FCGS_2$ technique. The horizontal axes indicate the nucleotides positions on the chromosome III of the *C.elegans* and the vertical axes indicate the frequency values.

These errors arise from the fact that intronic regions as recognised as intergenic ones, due to their important length. Here, we provided an efficient tool to easily recognize intronic regions; which will help in enhancing the gene prediction and the correction of available annotations.

## V. CONCLUDING REMARKS

In this paper, we turned our attention to introns study in the *C.elegans* genome by means of the complex Morlet wavelet transform. The intronic DNA is of a great interest given that is implicated in the dynamic regulation of gene as well as in the organism's evolution. In this work, we furnished two main contributions. The first contribution consisted on the DNA coding method, which is the Frequency Chaos Game Signal. The basic idea of this approach is to attribute numerical values to each of the DNA characters based on the Frequency Chaos Game Representation. The specificity of this kind of DNA representation relates to the genome statistical properties exploited in the coding procedure. Since it is the first time that such study is presented, the second contribution consists on applying the complex Morlet wavelet analysis to investigate the DNA characterization in the *C.elegans* independently of the sequences nature. In fact, studying bio-molecular sequences and exploring each of DNA regions with the same analysis tool formed a great challenge for several years. In our works, we have succeeded in exploring a multitude of biological information by adopting the complex Morlet wavelet transform.

For this paper, we have chosen results relating to the intronic regions in the *C.elegans*. The results elicited by the colour scalograms of the $FCGS_2$ have shown a very distinguished texture in introns of the *Y50D7A.4* gene. These textures are characterized by specific periodic patterns surrounding the frequency 1/6.5.Most of the intronic areas in the *C.elegans* genes are characterized by the periodicity 6.5 bp, but it is not the only one. Thus, as an extension to this work, we tend to study the different behaviours adopted in intron sequences. This will be very useful in establishing a genes database.